\documentclass[prl,showpacs,twocolumn,aps]{revtex4}
\usepackage{graphicx}
\newcommand{\be}{\begin{equation}}
\newcommand{\ee}{\end{equation}}
\newcommand{\bea}{\begin{eqnarray}}
\newcommand{\eea}{\end{eqnarray}}

\newcommand{\bwt}{\begin{widetext}}
\newcommand{\ewt}{\end{widetext}}
\begin{document}
\title{Pairing Fluctuations in The Attractive Hubbard Model in the Atomic Limit
}
\author{S. Verga$^1$, R.J. Gooding$^{2}$ , and F. Marsiglio$^{1}$}
\affiliation{
$^{1}$Department of Physics, University of Alberta, Edmonton, Alberta,
Canada, T6G~2J1
\\
$^{2}$ Department of Physics, Queen's University, Kingston,
Ontario, Canada, K7L~3N6}
\begin{abstract}
BCS theory accounts for the pairing instability in the weak coupling limit, but
fails to describe pairing fluctuations above $T_c$. One possibility for describing
these fluctuations in the dilute limit is the T-matrix approximation.
We critically examine various degrees of self-consistency in the
T-matrix formalism, along with a non-diagrammatic two-particle
self-consistent (TPSC) formulation, in the strong coupling regime, where
an exact solution is readily available. We find that one particular
degree of self-consistency is quite accurate, particularly at low temperature
as evidenced by examining both static and dynamic
properties. 
\end{abstract}

\date{\today}
\maketitle

In a recent essay titled, ``Brainwashed by Feynman'' \cite{anderson00},
P.W. Anderson
wrote about the different approaches to solving many-body problems.
His main point was that the solution to some
problems requires a leap beyond the ``tried and true procedures''
i.e. perturbation theory. He even argues that attempts to go beyond
perturbation theory, for example by ``summing all the diagrams'',
are more or less futile. We generally agree with these sentiments.

On the other hand, we would argue when such a creative step occurs,
as happened when BCS formulated their famous wavefunction for the
superconducting ground state \cite{bardeen57}, there 
nonetheless remains a desire to recast the solution in terms of diagrams.
In this particular case precisely this was accomplished shortly thereafter,
by Gor'kov \cite{gorkov58} and others. That this was indeed useful
is clear by the subsequent work that followed concerning generalizations
of BCS theory, such as Eliashberg theory \cite{eliashberg60}.
The work by Thouless \cite{thouless60} elucidated the nature of the
superconducting instability in terms of an infinite set of diagrams ---
the ladder sum in the particle-particle (Cooper) channel. 
This allowed one to study the possibility of a
superconducting transition away from the weak coupling limit for which
BCS was initially designed, and this indeed has occurred over the decades
since, as evidenced by, for example, Refs. \cite{kadanoff61,eagles69,
leggett80,nozieres85,janko97}. Many more references are cited in
Ref. \cite{gooding04}. All of this work is characterized by an
attempt to cast the superconducting instability in terms of a subset
of Feynman diagrams; they thus suffer from the ``Brainwashed by Feynman''
(BBF) attitude documented in Ref. \cite{anderson00}.  

We should note that in recent years some work attempts to tackle
the problem in a way which cannot be simply recast in terms of Feynman
diagrams \cite{vilk97,allen01,kyung01,pieri98}. This approach has
been reasonably successful as judged by comparison with (exact) Monte
Carlo results \cite{vilk97,allen01,kyung01}.

The purpose of this paper is to examine this problem once again, in
a limit for which an exact solution is readily available, i.e. the
atomic limit. This limit allows us to discriminate between candidate
theories; we find that one in particular emerges as very accurate,
particularly at low temperature.
For simplicity we will use the attractive Hubbard model
(AHM). The atomic limit is essentially in the strong coupling limit of
this model;
this is where the BBF approach should encounter the most difficulties.
Nonetheless we bravely
forge ahead, using the T-matrix
approach pioneered by Thouless and others \cite{kanamori63}. In the BBF
approach one of the questions that researchers have been addressing is:
what level of self-consistency (if any) most properly reproduces the
correct result for the superconducting instability over all regimes
(weak to strong coupling) ? 

In what follows we will formulate the
problem in a manner which avoids Hartree diagrams \cite{zlatic00}.
This minimizes the actual number of diagrams that need be considered,
and avoids the plethora of possible degrees of self-consistency. 
We also consider the two-particle self-consistent approach of
Vilk and Tremblay (VT) \cite{vilk97}, as 
implemented by Kyung et al. \cite{kyung01} for the AHM. The VT
approach apparently works best for intermediate coupling. Remarkably
it and one of the self-consistent T-matrix formulations both
become exact at zero temperature in this strong coupling limit.

Let us begin with the Hamiltonian. It is written \cite{zlatic00}
\bea
H & = & -t\sum_{i,j \atop \sigma} \bigl(\hat{c}_{i\sigma}^\dagger 
\hat{c}_{j\sigma} 
+ H.c.\bigr) - \mu^\prime \sum_{i,\sigma}\hat{n}_{i,\sigma}
\nonumber \\
 & & -|U| \sum_{i} \bigl(\hat{n}_{i,\uparrow} - n/2 \bigr) 
\bigl(\hat{n}_{i,\downarrow} - n/2 \bigr), 
\label{ham}
\eea
where the first term is the usual hopping term for an electron, the
second is the chemical potential term, and the third describes the
attractive interaction of strength $|U|$ between electrons on the
same site. Note that the mean field expectation value of the
electron density of a given spin species in the paramagnetic state,
$<\hat{n}_{i\sigma}> = n_\sigma = n/2$,
is subtracted from the electron density operator in the interaction
term. Similarly, the modified chemical potential, $\mu^\prime$, is
given in terms of the actual chemical potential, $\mu$, by the
relation $\mu^\prime = \mu + n|U|/2$. The use of this Hamiltonian
allows us to ignore all Hartree diagrams; they have been included 
automatically by using $\mu^\prime$ instead of $\mu$.

The exact solution to this Hamiltonian is easy to obtain in the atomic limit
(i.e. $|U| >> t$, so that the problem reduces to a single site problem) \cite{micnas95}:
We find
\be
G_\uparrow (z) = {1 - n_\downarrow \over z + \mu^\prime - |U| n_\downarrow}
+ {n_\downarrow \over z + \mu^\prime + |U| (1 - n_\downarrow)}, 
\label{green_ex}
\ee
and therefore, with the non-interacting limit, $G_0(z) = 1/(z + \mu^\prime)$,
as a reference state,
\be
\Sigma_\uparrow (z) = {|U|^2n_\downarrow (1 - n_\downarrow) \over
z + \mu^\prime + |U| (1 - 2 n_\downarrow)}.
\label{self_ex}
\ee
Note the simplicity of the exact solution.
Accounting for the Hartree term in the modified chemical potential,
the single electron propagator contains two poles, analogous to a lower
and upper Hubbard band in the repulsive model, with energies separated
by $|U|$.

The electron density $n$ is easy to obtain in terms of the modified chemical
potential $\mu^\prime$. Inverting this relation, one obtains
\be
\mu^\prime = -{|U| \over 2}\bigl( 1 - n \bigr) -T\ln{\bigl(a(n)/n\bigr)}
\label{muprime_ex}
\ee
Here the zero temperature result is given in the first term, and is quite
simple, and finite temperature corrections are given by the second term, with
\be
a(n) = (1-n)e^{-\beta |U|/2} + \sqrt{n(2-n) + 
(1-n)^2 e^{-\beta |U|}}. 
\nonumber
\ee
A side benefit of using $\mu^\prime$ 
is that the solutions are
spread out as a function of electron density, which would not be the case
if $\mu$ were used instead.
Some of these solutions are shown in Fig.~(1a). 
We also examine the two particle
correlations. One may derive
\be
<n_\uparrow n_\downarrow > = (1-n)N\bigl[ - 2\mu^\prime - |U|(1-n) \bigr],
\label{nupndo_ex}
\ee
where $N(x) \equiv 1/(\exp{(\beta x)} - 1)$ is the Bose function. With the
result of Eq.~(\ref{muprime_ex}) this `simplifies' to
\be
<n_\uparrow n_\downarrow > = {n^2 \over 2} 
{1 \over n + b(n)\exp{(-\beta |U|/2)}},
\label{nupndo_ex_aux}
\ee
and $b(n) = (1-n) e^{-\beta |U|/2} + \sqrt{n(2-n) + (1-n)^2 e^{-\beta |U|}}$. 
Again, as $T\rightarrow 0$ the result is 
quite simple and expected, $<n_\uparrow n_\downarrow > \rightarrow n/2$.

The T-matrix approximation to this problem results in a self energy of
the form \cite{kadanoff61,janko97,beach01}
\be
\Sigma(i\omega_m) = -|U|^2 {1 \over \beta} \sum_{\ell} {\chi_0(i\nu_{\ell}) \over
1 - |U| \chi_0(i\nu_{\ell})} G_c(-i\omega_m + i\nu_{\ell}),
\label{self_generic}
\ee
where $i\omega_m \equiv i\pi T(2m - 1)$ and $i\nu_{\ell} \equiv i2\pi T {\ell}$
are the Fermion and Boson Matsubara frequencies, respectively, and $m$ and
${\ell}$ are integers. The `bare' susceptibility $\chi_0(i\nu_n)$ is given
by
\be
\chi_0(i\nu_n)={1\over \beta}\sum_m G_a(i\omega_m)G_b(-i\omega_m+i\nu_n),
\label{chi0_generic}
\ee
where the subscripts $a$, $b$, and $c$ in the above two equations can
either be absent (to indicate that the fully interacting Green function
should be used) or can be set to `0', to indicate that the non-interacting
Green function is used. In the former case, one is required to use Dyson's
equation, $G(i\omega_m) = 1/(i\omega_m + \mu^\prime - \Sigma(i\omega_m))$
to self-consistently determine $G$ and $\Sigma$. While the literature
has some discussion of these various approximations, 
our purpose here is to examine them critically in the atomic limit.

\begin{figure}[tp]
\begin{center}
\includegraphics[height=3.2in,width=3.0in]{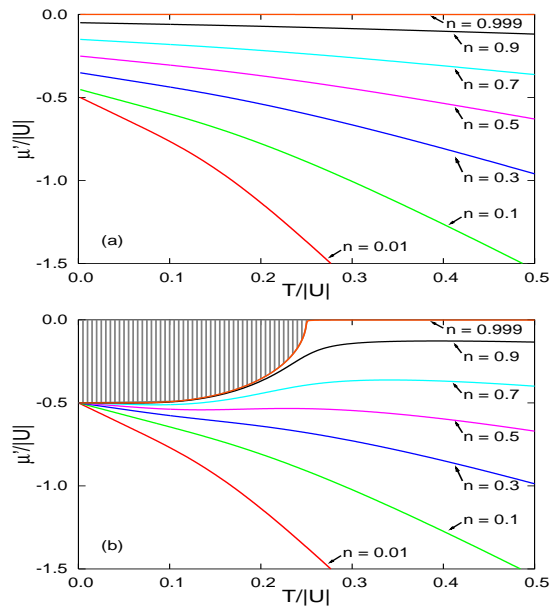}
\caption{(a) Exact solutions and (b) $(G_0G_0)G_0$ results for $\mu^\prime$
vs. temperature for different densities. In (b) the shaded region is
the so-called Thouless region. Note that it is absent in the exact 
solution.
}
\end{center}
\end{figure}

The non-self-consistent (NSC) approximation was first examined in two
dimensions for the AHM in Refs. \cite{schmitt-rink89,serene89}.
Fig.~(1b) shows the results in the atomic limit. Comparison with Fig.~(1a)
shows that this approximation is poor; indeed the same conclusion
occurs in two dimensions \cite{beach01}. A `Thouless region' in the $\mu^\prime - T$
plane occurs; this is where an instability would occur if $\mu^\prime$
was held fixed while the temperature is lowered; it is indicated in
Fig.~(1b) by the shaded region. The more physical procedure is to keep the
electron density fixed. Then $\mu^\prime$ adjusts so that a finite
temperature transition is avoided. As in higher dimensions, the
electrons form bound states, the Fermi sea is depleted, and a finite
temperature transition is avoided {\it for the wrong reason}, as inspection
of Fig.~(1a) shows.

We have examined all the possible variations for Eqs.(\ref{self_generic},
\ref{chi0_generic}) at low electron densities. The results for $\mu^\prime$ vs. $T$ at
a low electron density are shown in Fig.~(2a). In addition we have also
computed the result for the Vilk-Tremblay \cite{kyung01} theory. At very
low electron densities all approximations work reasonably well. At low
(but not too low) electron densities, two emerge as particularly accurate,
the $(GG_0)G_0$ T-matrix theory, and the Vilk-Tremblay theory. We use
a nomenclature for the T-matrix theory to correspond to $(G_aG_b)G_c$,
where the $a$, $b$, and $c$ refer to the labels in Eqs.(\ref{self_generic},
\ref{chi0_generic}).

\begin{figure}[tp]
\begin{center}
\includegraphics[height=3.2in,width=3.2in]{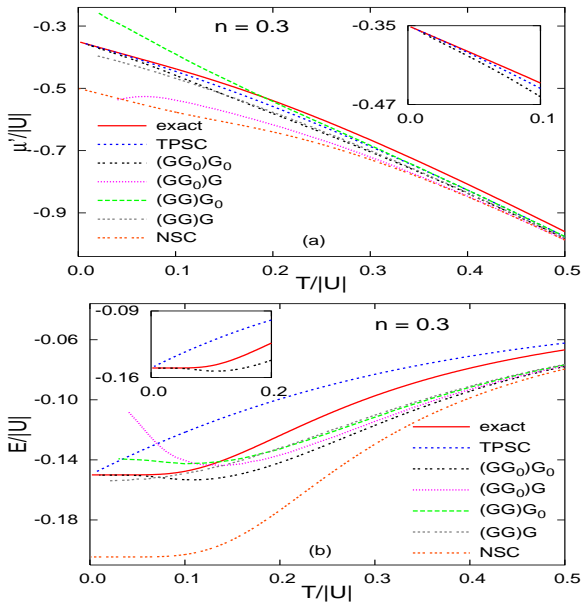}
\caption{A comparison of different approximations at n = 0.3
using (a) $\mu^\prime$, and (b) energy. Note that the VT result
(first iteration only) and the $(GG_0)G_0$ T-matrix approximation
are the most accurate; they both become exact as $T \rightarrow 0$.
The second iteration of the VT result (not shown) is poorer than 
the first. The insets focus on the TPSC and $(GG_0)G_0$
results at low temperatures; note in (b) how well the $(GG_0)G_0$
curve approaches the exact result over an extended temperature range. 
}
\end{center}
\end{figure}

Clearly, the Vilk-Tremblay result is most accurate. However, particularly
at low temperatures, it is evident that the $(GG_0)G_0$ T-matrix
approximation is also quite accurate. In fact, at lower electron densities
(not shown) the $(GG_0)G_0$ result becomes more accurate than the VT
result, and these two remain the front-runners for all electron densities.
Interestingly, both the VT and the $(GG_0)G_0$ results appear to
become exact at low temperatures. Indeed, in the $(GG_0)G_0$ formulation
the ansatz
\be
\Sigma(i\omega_m) = {\Delta^2 \over i\omega_m - \mu^\prime} 
\label{self-ansatz}
\ee
emerges if one considers only the term in Eq. (\ref{self_generic}) 
with ${\ell} = 0$. Then $\Delta^2 = T |U|^2 \chi_0(0)/(1 - |U| \chi_0(0))$.
Insertion of this self energy into the one electron Green function
allows one to evaluate the bare susceptibility in Eq. (\ref{chi0_generic})
at zero frequency. Earlier work \cite{beach01} indicated that any degree of
self-consistency in the 'bare' susceptibility drives the superconducting
transition (signalled by $1 = |U|\chi_0(0)$) to zero temperature.
Adopting this requirement in this case gives the parameter $\Delta^2$
in terms of $|U|$ and $\mu^\prime$. The number equation, $n = {2 \over \beta}
\sum_m G(i\omega_m)\exp{(i\omega_m0^+)}$ provides a second relation
between $\Delta^2$ and $\mu^\prime$; with these two equations all the zero
temperature properties can be obtained analytically, and, remarkably, they agree
with the exact solution. Thus, the suppression of the Thouless instability
to zero temperature is a feature in common with higher dimensional solutions.
For the VT theory, these results are for the first iteration \cite{kyung01}.
If one presses further with the second iteration, the results deteriorate,
consistent with the observation of these authors \cite{allen04}.

We can press further these comparisons, and in particular probe two-particle
correlations. An easy way to do this is by examining the energy per
lattice site $E$, which normally would be simply proportional to the
double occupancy in the atomic limit. However, because of the special 
form of the Hamiltonian, Eq. (\ref{ham}), the precise relation
is
\be
E = -|U| <n_\uparrow n_\downarrow > + |U| \bigl({n\over 2}\bigr)^2. 
\label{energy-exact}
\ee
We use the exact relation
\be
E = {1 \over \beta} \sum_m \Sigma(i\omega_m)G(i\omega_m)\exp{(i\omega_m0^+)},
\label{energy-tmatrix}
\ee
to obtain the energy at any temperature, in the T-matrix theories.
Fig.~(2b) shows the energy determined within the various approximation
schemes along with the exact and the VT result, for $n = 0.3$. 
Once again the VT and
$(GG_0)G_0$ results are exact at zero temperature. However, the
VT result deviates immediately for $T>0$, while the $(GG_0)G_0$ result
follows closely over some temperature range. That this latter result
is true follows also from the analytical work described above.

This agreement at $T=0$ (note that neither approximation is particularly accurate
at intermediate temperatures) raises the question of the origin of the effective 
potential, $|U_{pp}|$, in the VT theory. Within a conserving approxmiation,
a frequency-independent irreducible vertex, $|U_{pp}|$, is accompanied by
single electron propagators that include a simple Hartree-like renormalization
\cite{kyung01}. The $(GG_0)G_0$  
T-matrix approach, however, suggests that the origin of an effective interaction
is from the (partial) renormalization of the single electron propagator.
That is, we can construct a theory that appears similar to the VT theory, i.e. with
unrenormalized propagators everywhere in Eqs.(\ref{self_generic},\ref{chi0_generic}),
but with an {\it effective} interaction vertex. Eq.(\ref{self_generic}) suggests this will
be accomplished by
\be
|U|\chi_0(i\nu_n) \equiv |U_{eff}|\chi_{00}(i\nu_n),
\label{ueff}
\ee
where $\chi_{00}(i\nu_n) = {1 \over \beta}\sum_m G_0(i\omega_m)G_0(-i\omega_m+i\nu_n)$.
Note that this requires 
$|U_{eff}|$ to depend on Matsubara frequency, but, in the spirit of VT (and
many Parquet-like treatments of higher order corrections \cite{bickers04}),
we will use Eq.~(\ref{ueff}) at zero Matsubara frequency to define an effective
potential. Comparisons with $|U_{pp}|$ as obtained in the VT formalism show 
quantitative discrepancies, particularly as the temperature approaches zero. Thus,
the two theories differ more substantively than Eq.~(\ref{ueff}) would suggest. 

To probe further the analytical structure of the various approximations, we examine 
the spectral function ($A(\omega) \equiv -{\rm Im} 
G(\omega + i\delta)/\pi$). As already noted, the $(GG_0)G_0$ reproduces the
exact result at zero temperature. This is shown in Fig.~3, where the spectral function for
the $(GG_0)G_0$ has been obtained through Pad\'e approximants. 
The agreement with the exact result at low temperatures is remarkable.
On the other hand, the VT result in the first iteration gives a single pole, 
which is clearly inaccurate. The result obtained from the second iteration does
contain two poles. However, as shown, it is fairly inaccurate. Thus, even though 
both the VT theory and the  $(GG_0)G_0$ T-matrix approximation give exact results
for the chemical potential and the energy at low temperatures, only the latter 
fully reproduces the exact result as a function of frequency. Inspection of Fig.~2
shows that `improving' the degree of self-consistency deteriorates the agreement.
This would mean that at low temperatures some cancellation occurs between the fully
self-consistent T-matrix diagrams ($(GG)G$ theory) and the omitted vertex corrections.
This possibility remains to be shown, however.

\begin{figure}[tp]
\begin{center}
\includegraphics[height=1.6in,width=2.5in]{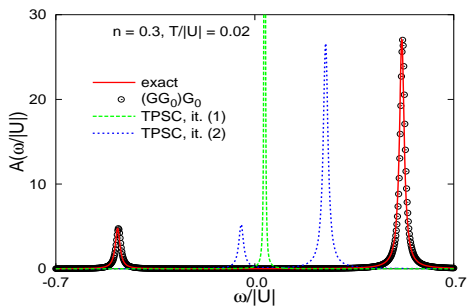}
\caption{ The spectral function for the $(GG_0)G_0$ T-matrix approximation (symbols),
compared with the exact result (solid curve). 
The remarkable agreement
shows that only partial self-consistency is a
requirement to reproduce the exact result.
The VT result (both first and second iterations) are poorer in
comparison. We have used a small artificial broadening, $\delta = 0.01|U|$ to
plot these results.
}
\end{center}
\end{figure}

In summary we have critically examined various approximations for the
attractive Hubbard model in the dilute and strong coupling limit. We
have used a variety of self-consistent T-matrix approaches along with the
two-particle self-consistent (VT) phenomenology. We found that minimal
self-consistency (the $(GG_0)G_0$ theory promoted in particular in Ref.
\cite{janko97}) and the VT calculation
both reproduced the exact result best, particularly at low
temperatures, where both become exact. Surprisingly, `improving' the
degree of self-consistency within a T-matrix approach leads to less
accurate results \cite{vilk97_comment}. Two particle correlations
(summarized in the total energy in this work) are faithfully reproduced
by the $(GG_0)G_0$ calculation for a range of low temperatures, and,
finally, the one electron spectral function is remarkably accurate, even at
nonzero temperature.
These results suggest that, at least for low electron densities, this
T-matrix formulation is useful for higher dimension calculations. It
would also be interesting to see how the various vertex corrections
provide partial cancellation to the omitted self-consistent 
contributions.

This work was supported in part by the Natural Sciences and Engineering
Research Council of Canada (NSERC), by ICORE (Alberta), and by the
Canadian Institute for Advanced Research (CIAR).
\bibliographystyle{prl}

\begin{thebibliography}{1}
\bibitem{vilk97_comment} Although arguments for why this might be expected
were advanced in Ref. \cite{vilk97}.
\end{thebibliography}

\end{document}